\documentclass{aastex}
\usepackage{psfig}

\def\chandra{{\it Chandra\/}}
\def\conx{{\it Constellation-X\/}}

\def\iue{{\it {\it IUE}\/}}
\def\rosat{{\it ROSAT\/}}

\def\xeus{{\it XEUS\/}}
\def\xmm{{\it XMM-Newton\/}}

\def\redshift{{\it z\/}}
\def\aox{$\alpha_{\rm ox}$}

\def\sdssfull{SDSSp~J104433.04--012502.2}
\def\sdss{SDSS~1044--0125}

\def\ltsima{$\; \buildrel < \over \sim \;$}
\def\simlt{\lower.5ex\hbox{\ltsima}}
\def\gtsima{$\; \buildrel > \over \sim \;$}
\def\simgt{\lower.5ex\hbox{\gtsima}}

\slugcomment{Accepted for publication in The Astronomical Journal}

\shorttitle{AN XMM-NEWTON DETECTION OF \sdssfull}

\shortauthors{BRANDT ET~AL.}

\begin{document}

\title{An XMM-Newton detection of the \redshift\ = 5.80 X-ray weak quasar \sdssfull} 


\author{
W.N.~Brandt,\altaffilmark{1} 
M.~Guainazzi,\altaffilmark{2} 
S.~Kaspi,\altaffilmark{1}  
X.~Fan,\altaffilmark{3}  
D.P.~Schneider,\altaffilmark{1}  
Michael A.~Strauss,\altaffilmark{3}  
J.~Clavel,\altaffilmark{2} 
and
J.E.~Gunn\altaffilmark{3}}
\altaffiltext{1}{Department of Astronomy \& Astrophysics, 525 Davey
Laboratory, The Pennsylvania State University, University Park, PA, 16802.}
\altaffiltext{2}{\xmm\ Science Operations Center, VILSPA-ESA, Apartado 50727, 28080 Madrid, Spain.}
\altaffiltext{3}{Princeton University Observatory, Princeton, NJ, 08544.}

\begin{abstract}
We report on an \xmm\ observation of the most distant known quasar, \sdssfull,  
at $z=5.80$. We have detected this quasar with high significance in the
rest-frame 3.4--13.6~keV band, making it the most distant cosmic object
detected in X-rays; $32\pm 9$ counts were collected. \sdssfull\
is notably X-ray weak relative to other luminous, optically selected quasars, 
with $\alpha_{\rm ox}=-1.91\pm 0.05$ and a 3.4--13.6~keV luminosity of 
$\approx 1.8\times 10^{44}$~erg~s$^{-1}$. The most likely reason for its X-ray 
weakness is heavy absorption with $N_{\rm H}\simgt 10^{24}$~cm$^{-2}$, as 
is seen in some Broad Absorption Line quasars and related objects; 
we discuss this and other possibilities. High-quality spectroscopy 
from 0.95--1.10~$\mu$m to search for blueshifted C~{\sc iv} 
absorption may elucidate the origin of the X-ray weakness.  
\end{abstract}



\section{Introduction}
\label{intro}

X-ray observations of quasars at the highest redshifts should reveal the 
physical conditions in the immediate vicinities of their central black holes.
Measurement of the intrinsic X-ray continuum's shape, normalization relative to the 
rest of the spectral energy distribution, and variability can provide
information on the inner accretion disk and its corona, and thus ultimately
about how the black hole is being fed. There are already some
reports that the X-ray continuum shapes of quasars may evolve with redshift
at $z\simlt 2.5$ (e.g., Blair et~al. 2000; Vignali et~al. 1999); these 
intriguing results require further study and extension to the highest 
redshifts. 
In addition, X-ray absorption measurements can probe material in the
environments of high redshift quasars. For example, the fraction 
of radio-loud quasars (RLQs) with heavy intrinsic and/or associated 
X-ray absorption (i.e., $N_{\rm H}\simgt 5\times 10^{22}$~cm$^{-2}$) 
appears to rise with redshift (e.g., Fiore et~al. 1998; Elvis et~al. 1998; 
Reeves \& Turner 2000). The absorbing gas in RLQs may be 
circumnuclear, located in the host galaxy, or entrained by the 
radio jets. 

At $z>4$ our knowledge about the X-ray properties of quasars is unfortunately
quite limited (e.g., Kaspi, Brandt \& Schneider 2000 and references therein).
To date it has only been possible to obtain X-ray spectra for four $z>4$
quasars. All four are radio-loud blazars in which the X-ray emission is probably 
dominated by jets; three of these appear to have X-ray absorption, as would 
be expected from the trend mentioned above (Moran \& Helfand 1997; 
Boller et~al. 2000; Fabian et~al. 2000; Yuan et~al. 2000). The more 
representative radio-quiet quasars (RQQs) have been too X-ray faint to
allow spectroscopy to date; indeed, only 7 optically selected quasars have even
been detected in X-rays at $z>4$. At present, measurements of the X-ray spectral
shapes of RQQs are scarce and of limited statistical quality above 
$z\approx 2.5$ (e.g., Bechtold et~al. 1994; Vignali et~al. 1999). 

We have started a project to determine the X-ray properties of the highest 
redshift quasars using the new generation of X-ray observatories. Our main
focus is on quasars with $z>4.8$ discovered by the Sloan Digital Sky Survey
(SDSS; see York et~al. 2000). The SDSS multicolor selection method
provides a uniform sample of quasars that have been consistently selected
in a well-defined manner, and more than 100 SDSS quasars at $z>5$ are
expected over the next five years (Fan 1999; Schneider 1999). 
After the recent discovery of \sdssfull\ 
(hereafter \sdss; Fan et~al. 2000), an optically bright RQQ
(AB$_{1280}=19.28$; $R=f_{\rm 6~cm}/f_{4400~\mbox{\footnotesize \AA}}<11$) 
at $z=5.80$, we proposed an \xmm\ Target of Opportunity 
observation of this object. From the Eddington limit, \sdss\ appears to 
contain a $\simgt 1.6\times 10^9$~$M_\odot$ black hole formed within a 
billion years of the Big Bang (we adopt $H_0=70$~km~s$^{-1}$~Mpc$^{-1}$ 
and $q_0=0.5$ throughout; note this cosmology differs from those 
of Fan et~al. 2000). The quasar's luminosity density at 2500~\AA\ is
$l_{2500~\mbox{\footnotesize \AA}}=2.4\times 10^{31}$~erg~s$^{-1}$~Hz$^{-1}$,
about twice that of 3C~273 (we adopt an optical 
continuum slope of $\alpha_{\rm o}=-0.79$
for \sdss\ throughout; see Schneider, Schmidt \& Gunn 1991 
and Fan et~al. 2001). Typical X-ray to optical flux ratios for optically 
selected $z>4$ quasars of similar luminosity (Kaspi et~al. 2000) 
suggested that \xmm\ might be able to obtain a spectrum of this object 
in a reasonable integration time. This paper describes the 
analysis and interpretation of the \xmm\ data. 


\section{Observations and data analysis}
\label{obser}

\subsection{Basic analysis}

\sdss\ was observed by \xmm\footnote{See the \xmm\ Users' 
Handbook at http://xmm.vilspa.esa.es/user/uhb/xmm\_uhb.html.
Also see Jansen (1999).}
starting on 2000 May 28; the observation was continuous,
and the quasar was placed at the aim point. 
Here we will focus on the results from the
European Photon Imaging Camera (EPIC) 
positive-negative junction
(p-n; 40.0~ks exposure) 
and 
Metal Oxide Semiconductor (MOS; 32.5~ks exposures each) 
detectors; the thin optical blocking filters were used with 
these detectors. The data were processed with the 
\xmm\ Science Analysis System (SAS) Version~4.1 
pipeline.\footnote{See the \xmm\ SAS documentation 
at http://xmm.vilspa.esa.es/sas/. Also see Page (1998).}
We have screened the resulting events using only 
event patterns 0--3/0--12 for the p-n/MOS
(see \S21 of Gondoin 2000), and we have removed 
7.4/9.7~ks of p-n/MOS data where the particle 
background showed flares. 

The screened events from the above reduction were analyzed
using the SAS and the \chandra\ Interactive Analysis of 
Observations (CIAO) software package.\footnote{See the CIAO 
documentation at http://asc.harvard.edu/ciao/. 
Also see McDowell, Noble \& Elvis (1998) and Elvis et~al., in preparation.}
We made images for each of the detectors
in the 0.5--2.0 and 2--7~keV bands; these bands provide
optimal signal-to-noise ratio for the study of faint sources
and cover the 3.4--48~keV rest-frame continuum. 
We improved the astrometry of these images
using the MOS-detected H~{\sc ii}/Seyfert galaxy Mrk~1261
(e.g., Osterbrock \& Phillips 1977) as well as 13 sources 
from the Palomar Optical Sky Survey; both a rotation and
a shift were required. We consider our absolute astrometry 
to be accurate to within $\approx 2^{\prime\prime}$.

\subsection{Source detection}

We have searched the images for sources using several algorithms, 
and we detect a 0.5--2.0~keV p-n source coincident with the  
optical position of \sdss\ with high statistical significance
(see Figure~1). For example, {\sc wavdetect} (Freeman et~al. 2000) 
finds this source when run with a significance threshold of 
$7.5\times 10^{-8}$; given the number of X-ray positions searched
that are consistent with the optical position, the probability
that the X-ray source is only a statistical fluctuation is 
$\simlt 10^{-5}$. 
Similarly, a detection algorithm created from the SAS tools finds 
this source when run with a minimum detection likelihood of 10
(see Cruddace et~al. 1988), corresponding to a false detection 
probability of $\approx 4.5\times 10^{-5}$. For background 
estimation, the SAS algorithm used a two-dimensional spline fit 
to the image with the sources removed. 
Our best position for the X-ray source is 
$\alpha_{2000}=$ 10$^{\rm h}$ 44$^{\rm m}$ 33.0$^{\rm s}$,
$\delta_{2000}=$ $-01^\circ$24$^{\prime}$58.0$^{\prime\prime}$
with a $1\sigma$ error circle radius of $5^{\prime\prime}$
(the positional error here is larger than the absolute astrometry 
error quoted above due to the small number of photons associated
with the X-ray source; this limits the precision of determination 
of the source's centroid). The X-ray source is $4.2^{\prime\prime}$ 
from \sdss. 
The quasar is not detected with high significance in the 2--7~keV p-n 
image (although manual inspection shows a hint of a photon excess 
at the quasar's position) or in any of the MOS images. The 0.5--2.0~keV 
MOS nondetections are not surprising, given their lower sensitivity
and the measured p-n flux (see below). 

We consider source confusion problems to be unlikely (compare with 
\S4.1 of Schmidt et~al. 1998). For example, the 0.5--2.0~keV 
$\log N$--$\log S$ relation of Hasinger et~al. (1998) indicates that the
probability of having a confusing $\geq 10^{-15}$~erg~cm$^{-2}$~s$^{-1}$
(see \S2.3) X-ray source within $5^{\prime\prime}$ of \sdss\ is 
$\approx 6\times 10^{-3}$. In addition, there are no optical or 
near-infrared sources other than \sdss\ inside the 
$5^{\prime\prime}$-radius X-ray error circle down to 
$i^\prime\approx 22$, 
$z^\prime\approx 20$ and
$K^\prime\approx 22.3$
(see \S2 of Fan et~al. 2000). 
For that matter, there are no such sources within 
$20^{\prime\prime}$ of \sdss. 

\subsection{Source parameterization}

We have performed aperture photometry to determine the number of
0.5--2.0~keV counts detected from \sdss. We have used a 
circular source cell of radius $14^{\prime\prime}$; this
encircles $\approx 65$\% of the 0.5--2.0~keV energy. 
We detect $31.7\pm 8.5$ net counts (corrected for the encircled 
energy fraction). The resulting 0.5--2.0~keV net count rate is 
$(9.72\pm 2.61)\times 10^{-4}$~count~s$^{-1}$. 
We have also performed photometry with a point-spread-function 
fitting technique, and the results are statistically consistent
with our aperture photometry. 
We have checked the distribution of photon arrival times, and
they appear to be spread fairly uniformly throughout the observation;
this argues against some brief, transient event (e.g., a
cosmic ray strike or transient hot pixel) having produced a
spurious source. 
We do not detect enough counts from \sdss\ to allow spectral fitting, 
so we have calculated its flux and luminosity adopting a power-law
model with a photon index of $\Gamma=2$ and a Galactic neutral hydrogen 
column density of $4.6\times 10^{20}$~cm$^{-2}$ (Stark et~al. 1992). 
We have chosen $\Gamma=2$ based upon the photon indices measured
for other luminous RQQs above 2~keV in the rest frame
(e.g., Reeves \& Turner 2000 find a typical range of 
$\Gamma=$~1.7--2.3), and our spectral model is consistent with 
our 0.5--2.0~keV detection and 2--7~keV upper limit 
(see below). Using the Portable, Interactive, Multi-Mission
Simulator ({\sc pimms}) Version~3.0 software 
(Mukai 2000), we find a 0.5--2.0~keV absorbed flux of 
$(1.06\pm 0.28)\times 10^{-15}$~erg~cm$^{-2}$~s$^{-1}$; 
the 0.5--2.0~keV flux corrected for Galactic absorption is 
$(1.22\pm 0.33)\times 10^{-15}$~erg~cm$^{-2}$~s$^{-1}$
(we have verified that the use of {\sc pimms} is appropriate
given our event pattern screening in \S2.1). 
With a luminosity distance of 35900~Mpc, the 3.4--13.6~keV 
rest-frame luminosity is $(1.78\pm 0.48)\times 10^{44}$~erg~s$^{-1}$. 
We have confirmed that our detected flux is reasonable by calculating
the 0.5--2.0~keV $\log N$--$\log S$ relation for our field and 
comparing it with the $\log N$--$\log S$ of Hasinger et~al. (1998); 
such `$\log N$--$\log S$ fitting' allows an independent 
calibration of the flux scale.

Figure~2 compares the observed X-ray flux of \sdss\ 
(corrected for Galactic absorption) to that of other
$z>4$ quasars with similar AB$_{1450}$ magnitudes and 
optical luminosities. \sdss\ is X-ray fainter by a factor
of $\simgt 10$ relative to its peers; we shall discuss 
this further below. 

We have set an upper limit on the number of counts from 
\sdss\ in the 2--7~keV band again using a $14^{\prime\prime}$-radius
aperture; our aperture encircles $\approx 60$\% of the 2--7~keV
energy. Following Kraft, Burrows \& Nousek (1991), 
we derive a 95\% confidence upper limit of 12.3 counts (again corrected
for the encircled energy fraction).  
Using the same spectral model as was used two paragraphs above, 
we find a 2--7~keV absorbed flux of 
$<2.0\times 10^{-15}$~erg~cm$^{-2}$~s$^{-1}$ 
and a 13.6--48~keV rest-frame luminosity of 
$<3.0\times 10^{44}$~erg~s$^{-1}$. 
Given the 3.4--13.6~keV luminosity and spectral model from
two paragraphs above, we would expect a 13.6--48~keV
luminosity of $(1.60\pm 0.43)\times 10^{44}$~erg~s$^{-1}$,
below our upper limit; hence the lack of a rest-frame 
13.6--48~keV detection is not surprising. 

We have calculated \aox, the slope of a nominal power law between 
2500~\AA\ and 2~keV in the rest frame. We again adopt $\Gamma=2$ for 
the X-ray continuum, and we use an optical power-law slope of
$\alpha_{\rm o}=-0.79$ (see \S1) to calculate the 2500~\AA\ 
flux density from the AB$_{1280}$ magnitude of 19.28
(corresponding to AB$_{1450}=19.17$). 
We find $\alpha_{\rm ox}=-1.91\pm 0.05$. 
The derived value of $\alpha_{\rm ox}$ is not highly sensitive
to the X-ray and optical continuum shapes adopted, as expected. 
For example, if we instead use $\Gamma=2.3$ ($\Gamma=1.7$) and
recalculate $\alpha_{\rm ox}$ in a consistent manner, we obtain
$\alpha_{\rm ox}=-1.86\pm 0.05$
($\alpha_{\rm ox}=-1.97\pm 0.05$).

\section{Discussion}

\subsection{The X-ray weakness of \sdss} 

With this \xmm\ observation, \sdss\ surpasses GB~1428+4217 
($z=4.72$; Fabian et~al. 1997) as the highest redshift cosmic object 
detected in X-rays. The most notable X-ray characteristic of
\sdss\ is its faintness in this band (see Figure~2); we had 
expected to obtain an X-ray spectrum for \sdss\ but were foiled
by the surprisingly low X-ray flux. 
X-ray weak quasars are known to exist at low redshift where they 
comprise $\approx 10$\% of the optically selected quasar population
(e.g., Brandt, Laor \& Wills 2000, hereafter BLW). There is
increasing evidence that the majority of such objects are X-ray
weak because they suffer from intrinsic absorption: 
$\simgt 80$\% of optically selected, X-ray weak quasars show 
strong (4--80~\AA\ equivalent width), blueshifted
C~{\sc iv} absorption in the ultraviolet (BLW), and direct
X-ray spectral fitting has revealed absorbed (but otherwise
normal) X-ray continua in several X-ray weak quasars
(e.g., Gallagher et~al. 2000 and references therein). 
Broad Absorption Line quasars (hereafter BAL~QSOs), for example, 
are notoriously X-ray weak (e.g., Green \& Mathur 1996; 
Gallagher et~al. 1999). The absorbing column densities associated 
with X-ray weak quasars are often $N_{\rm H}\simgt 10^{23}$~cm$^{-2}$ 
and in some cases appear to be `Compton thick' 
(i.e., $N_{\rm H}\geq 1.5\times 10^{24}$~cm$^{-2}$). 

\subsubsection{Comparison with the Bright Quasar Survey} 

Sharper insight into the X-ray weakness of \sdss\ can be 
gained by comparing it with the $z<0.5$ Bright Quasar 
Survey (BQS; Schmidt \& Green 1983). This sample is fairly 
large and has been extensively studied at many wavelengths;
the high-quality constraints on X-ray and ultraviolet 
absorption are particularly relevant here. 

If \sdss\ is not absorbed in the X-ray band, its X-ray weakness 
is extreme compared to other unabsorbed RQQs. In Figure~3, we 
show a histogram of \aox\ for the 53 luminous ($M_{\rm V}<-22.27$;
we use the absolute magnitudes of Boroson \& Green 1992 converted
to our cosmology) BQS RQQs. The \aox\ values used are from BLW 
(calculated following Footnote~1 of that paper). 
We have shaded the histogram for 7 of these RQQs, since there 
is strong evidence that their \aox\ values have been depressed 
by intrinsic X-ray absorption. The shaded RQQs are the 
BAL~QSOs 0043+039, 1001+054, 1700+518 and 2112+059 
and the absorbed RQQs 1126--041, 1351+640 and 1411+442
(see Brinkmann et~al. 1999; Gallagher et~al. 1999; 
Wang et~al. 1999; BLW; Gallagher et~al. 2000).
The mean \aox\ and \aox\ standard deviation for the unshaded
BQS RQQs in Figure~3 are $-1.56$ and 0.14, respectively. \sdss, 
with $\alpha_{\rm ox}=-1.91\pm 0.05$, has a larger negative value 
of \aox\ than {\it all but one\/} of the unshaded BQS RQQs in 
Figure~3. 

The three unshaded BQS RQQs in Figure~3 with the lowest \aox\ 
values are 1259+593, 1552+085 and 2214+139. As we shall now
describe, there is some evidence that at least two of these 
have had their \aox\ values depressed by absorption (we have not 
shaded them in Figure~3 since the data are limited, and we wanted 
to be appropriately conservative).
2214+139 ($\alpha_{\rm ox}=-2.06$) has been observed by \rosat, 
but unfortunately the exposure time was fairly short. While 
detailed spectral fitting is not possible due to the small number 
of counts collected, its flat X-ray continuum shape 
($\Gamma=-0.07\pm 0.94$ when only Galactic absorption is assumed) 
strongly suggests intrinsic absorption (Rachen, Mannheim \& Biermann 1996). 
1552+085 ($\alpha_{\rm ox}=-1.80$) is one of the most highly polarized 
BQS quasars in the optical (1.9\%; Berriman et~al. 1990); a significant 
fractional contribution from scattered, polarized light most readily occurs 
when the direct continuum is absorbed, and most optically selected RQQs 
with polarization percentages $>1.5$\% show X-ray absorption 
(e.g., BLW and references therein). Furthermore, based on
International Ultraviolet Explorer (\iue) data, 
Turnshek et~al. (1997) argue that 1552+085 is a BAL~QSO, and
the notably weak [O~{\sc iii}] emission of 1552+085 is also 
often found in absorbed RQQs (e.g., Turnshek et~al. 1997; BLW). 
1259+593 ($\alpha_{\rm ox}<-1.79$) has the poorest X-ray coverage
of any of the $z<0.5$ BQS RQQs; the only limit available is a
fairly loose one from the \rosat\ All-Sky Survey. We are not aware 
of any multiwavelength observations which either strongly suggest 
or argue against intrinsic absorption in 1259+593. 

\subsubsection{Comparison with Other Quasar Samples} 

The X-ray weakness of \sdss\ is also apparent when it is compared 
with other quasar samples having higher median luminosities than
the BQS. Such comparisons are germane since \aox\ appears to generally
decrease (i.e., move towards larger negative values) with increasing 
luminosity (e.g., Avni, Worrall \& Morgan 1995 and references
therein); for our cosmology, \sdss\ is $\approx 40$ times more 
luminous than the typical BQS RQQ shown in Figure~3. 
Unfortunately, these higher luminosity samples do not 
have as complete multiwavelength coverage as the BQS, so it 
is more difficult to account for the significant effects of 
absorption that plague many \aox\ studies (see BLW and \S3.1.1
for discussion). In addition, precision comparisons with a sample
matched in luminosity are hindered because, for a reasonable
range of $q_0=$~0.1--0.5, the luminosity of \sdss\ at $z=5.80$ 
varies by a factor of $\approx 3.5$; all well-studied comparison 
samples have substantially lower median redshifts, and thus their 
luminosities depend much less on $q_0$. 

In Figure~4, we compare the \aox\ value of \sdss\ with \aox\ 
values derived for other RQQs found in optically selected 
surveys. In particular, the Large Bright Quasar Survey 
(LBQS; Hewett, Foltz \& Chaffee 1995) RQQs provide
a well-defined, optically selected comparison sample with a
significantly higher median luminosity than the BQS; the X-ray 
properties of these RQQs were studied by Green et~al. (1995) using 
an X-ray counts stacking technique. Comparison of \sdss\ with the 
highest luminosity LBQS data point in Figure~4 affirms its X-ray 
weakness. Furthermore, it is important to note that the LBQS data 
points include some absorbed RQQs (e.g., BAL~QSOs; see \S6 of
Green et~al. 1995). Removal of the absorbed RQQs would shift the LBQS 
data points towards less negative values of \aox, increasing their
separation from \sdss. The magnitude of this shift is expected to be 
$<0.1$ (P.J. Green 2000, private communication), largely due to the 
limited sensitivity \rosat\ All-Sky Survey observations used by 
Green et~al. (1995). 

We have used the weighted orthogonal regression for the RQQs in
Table~2 of Green et~al. (1995) to predict the expected \aox\
value for \sdss\ given its luminosity density at 2500~\AA\
(see \S1).\footnote{Revised values for the slope and intercept
in Table~2 of Green et~al. (1995) are $0.157$ and $-3.267$, 
respectively (P.J. Green 2000, private communication).} We predict 
$\alpha_{\rm ox}=-1.71$, in good agreement with expectations 
from Figure~4. 

Figure~4 also includes the optically selected $z>4$ RQQs studied 
by Kaspi et~al. (2000). We have re-calculated the 2500~\AA\ luminosity 
densities and \aox\ values from this paper using the optical 
continuum slope measurements from Schneider et~al. (1991) or the 
mean value of $\alpha_{\rm o}=-0.79$ from Fan et~al. (2001). The
\aox\ value for \sdss\ is substantially below the range 
found for these quasars. 
Finally, we note that \sdss\ has a larger negative \aox\ value 
than all but one of the RQQs studied by Bechtold et~al. (1994) 
even though these RQQs are significantly more luminous in the
rest-frame optical band than \sdss. 

In conclusion, if \sdss\ does not suffer from intrinsic and/or
associated X-ray absorption, then its anomalously weak X-ray continuum 
marks it as an unusual quasar. 

\subsection{Intrinsic and/or associated X-ray absorption?} 

The results of the above discussion suggest that heavy absorption 
may be affecting the X-ray spectrum of \sdss, and heavy absorption 
(specifically, a low-energy cutoff) is indeed commonly found in the 
X-ray spectra of $z\simgt 2.5$ RLQs (see the references in \S1). 
Is it possible that X-ray absorption generally increases with 
redshift in RQQs at $z\simgt 2.5$? In this case, the 
large negative \aox\ value for \sdss\ might be the result of this 
trend in extremis, and a few RQQs at $z\approx 2.5$ 
do appear to show significant X-ray absorption (see Figure~9 of
Reeves \& Turner 2000). While we consider this to be an enticing 
possibility requiring further investigation, we note that the 
\aox\ values for the $z>4$ RQQs in 
Kaspi et~al. (2000) do not appear to support it. 
Thus, the most likely interpretation at present appears to be 
that \sdss\ is just one of the minority of optically selected 
RQQs that have intrinsic and/or associated absorption depressing 
their X-ray continua (rather than a representative example of a
widespread, general RQQ trend). For example, \sdss\ may be a 
BAL~QSO or mini-BAL~QSO.  
Finally, we note that because the high redshift of \sdss\ has 
provided us with access to penetrating, high-energy X-rays, any 
absorber must be very thick with $N_{\rm H}\simgt 10^{24}$~cm$^{-2}$
(perhaps with partial covering of the X-ray continuum source to 
explain the photons detected at $\simlt 8$~keV). 
 
The integrated column density of the intergalactic medium out to 
$z=5.80$ is almost certainly too small to produce the required X-ray
absorption, especially given the intergalactic medium's low 
metallicity (e.g., Miralda-Escud\'e 2000). 


\subsection{An intrinsically weak X-ray continuum?} 

We must also admit the possibility that \sdss\ has an 
intrinsically weak X-ray continuum; such objects are rare
at low redshift, but some may exist (e.g., BLW; Gallagher et~al. 2000). 
Speculating for a moment, \sdss\ could be a `precursor quasar' 
where the relativistically deep potential well needed for efficient
X-ray production is still in the process of formation 
(see \S3.3 of Haehnelt \& Rees 1993); at ultraviolet and longer 
wavelengths, such contracting precursors may be able to sustain 
quasar-level luminosities for $\sim 10^7$ years. 
Alternatively, the black hole in \sdss\ could be accreting at
a highly super-Eddington rate where `trapping radius' effects 
cause the X-rays created in the inner region of the accretion 
flow to be dragged back into the hole (e.g., Begelman 1978; Rees 1978);
such an accretion flow has a luminosity of about the 
Eddington limit even though the mass accretion rate is
super-Eddington. To first order, radiation trapping would truncate 
the spectrum above a critical frequency depending on the trapping 
radius; it could therefore lead to weak X-ray emission. 
Super-Eddington accretion would help to explain the formation of 
such a massive black hole in less than a billion years (see \S1). 

Even if \sdss\ is intrinsically X-ray weak, this should not 
dramatically change the bolometric luminosity estimated in 
\S4.2 of Fan et~al. (2000). In luminous RQQs, X-rays typically 
contribute $\simlt 10$\% of the bolometric luminosity (e.g., \S6.1
of Elvis et~al. 1994). 

\subsection{Future studies} 

Searches for intrinsic and/or associated 
absorption lines in the rest-frame ultraviolet 
spectrum of \sdss\ are now of key importance for further understanding 
the origin of its X-ray weakness. Specifically, most of the X-ray weak 
quasars at low redshift that suffer from X-ray absorption also 
show significant blueshifted ultraviolet absorption; BQS RQQs 
with comparable \aox\ values to \sdss\ typically have C~{\sc iv} 
absorption equivalent widths of 4--80~\AA\ (BLW). Due to
the high redshift of \sdss, it does not yet have spectral coverage
of C~{\sc iv} (this line has an observed-frame wavelength of 1.05~$\mu$m).
Absorption lines from N~{\sc v} and Si~{\sc iv} are frequently 
seen in objects with strong C~{\sc iv} absorption, and these lines
are covered by the spectrum shown in Figure~2 of Fan et~al. (2000). 
One can make a case for a moderate strength Si~{\sc iv} absorption 
line blueshifted by $\approx 6000$~km~s$^{-1}$ from the emission line
(with a maximum rest-frame equivalent width of $\approx 6$~\AA), but 
this could simply be a dip in the continuum. Any N~{\sc v} absorption
at this velocity would be difficult to disentangle from the Ly$\alpha$ 
forest; there is none apparent at lower velocities. 
High-quality spectroscopy of the C~{\sc iv} line is required to 
settle this issue. 
Rest-frame ultraviolet absorption studies may well be the only way to
better understand the X-ray weakness of \sdss\ in the short term. 
Since the low X-ray flux of \sdss\ puts it below the
practical spectroscopy limit for even \conx\ (see Figure~2), directly
addressing the origin of its X-ray weakness with X-ray 
spectroscopy will almost certainly be impossible until 
\xeus\ or a comparable mission begins operation. 

More generally, further X-ray observations of the highest redshift
quasars are now needed to determine the distribution of \aox\ for
these objects. \chandra\ and \xmm\ studies of additional 
quasars found by the SDSS will soon address this issue. 


\acknowledgments
This paper is based on observations obtained with \xmm, an ESA science 
mission with instruments and contributions directly funded by 
ESA Member States and the USA (NASA).
We thank all the members of the \xmm\ and SDSS teams for their enormous efforts, and 
we thank F.A. Jansen for kindly allocating the time for this observation. We thank 
P.E. Freeman, S.C. Gallagher, P.J. Green, F.A. Jansen, 
D. Lumb, P. M\'esz\'aros, S. Sigurdsson, K.A. Weaver,
and an anonymous referee for helpful discussions. 
We gratefully acknowledge the financial support of 
NSF CAREER grant AST-9983783 and the Alfred P. Sloan Foundation (WNB),
NASA LTSA grant NAG5-8107 (SK),
NSF grant AST-9900703 (DPS), and  
NSF grant AST-9616901, Research Corporation, and a 
Porter O. Jacobus Fellowship (XF, MAS). 



\newpage


\begin{figure}[t!]
\hbox{
\hspace*{0.6 in} \psfig{figure=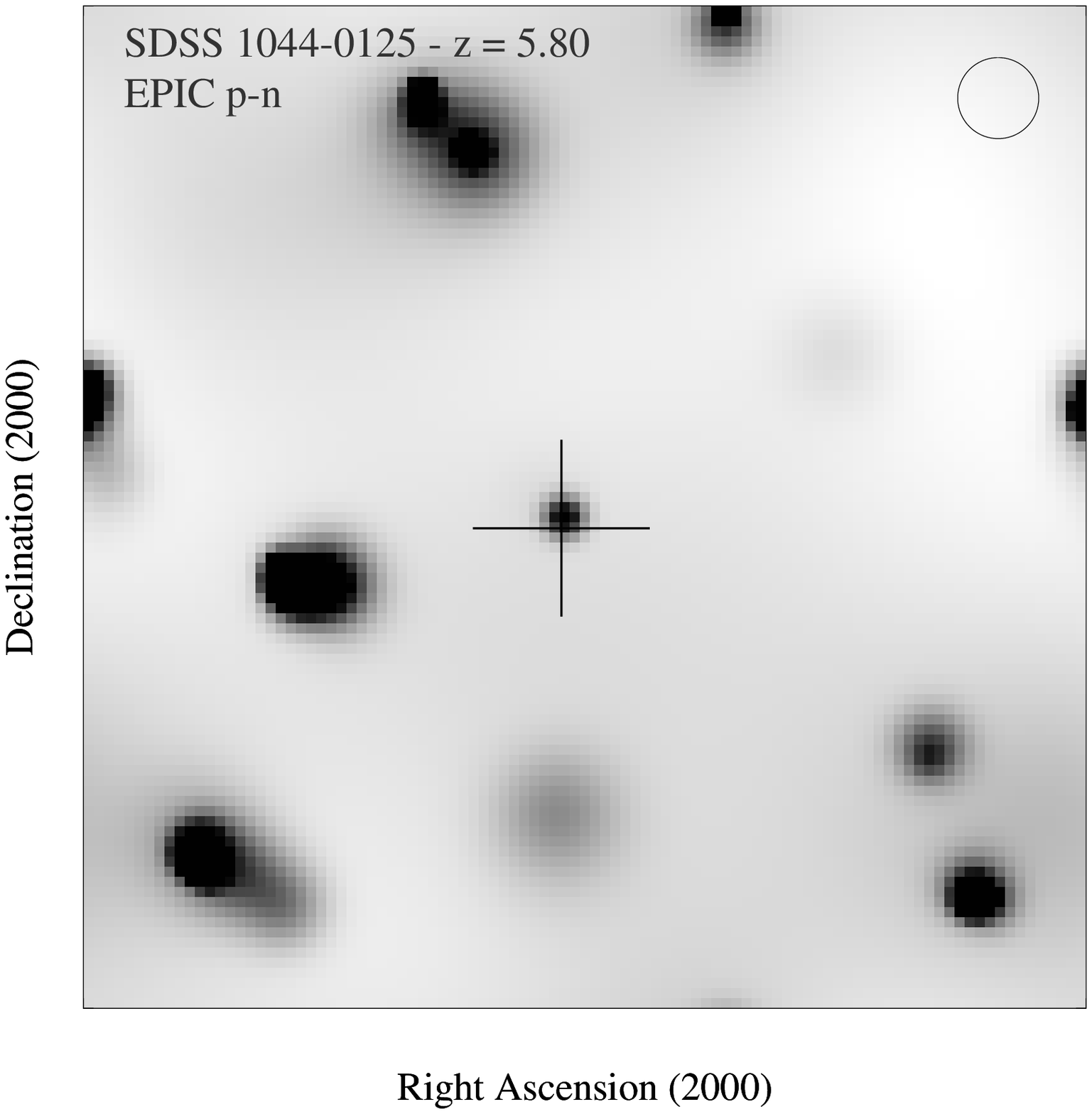,height=5.0truein,width=5.0truein,angle=0.}
}
\caption{EPIC p-n image of \sdss\ from 0.5--2.0~keV. The image has been 
adaptively smoothed at the $2.8\sigma$ level using the algorithm of 
Ebeling, White \& Rangarajan (2000). North is up, and East is to the
left. The image is $5.9^\prime$ on a side. The gray scale is linear. 
The cross marks the optical position of \sdss. The circle in the 
upper-right shows the size of the $14^{\prime\prime}$-radius cell
that was used for source count extraction; this encircles
$\approx 65$\% of the 0.5--2.0~keV energy.}
\end{figure}


\begin{figure}[t!]
\hbox{
\hspace*{0.6 in} \psfig{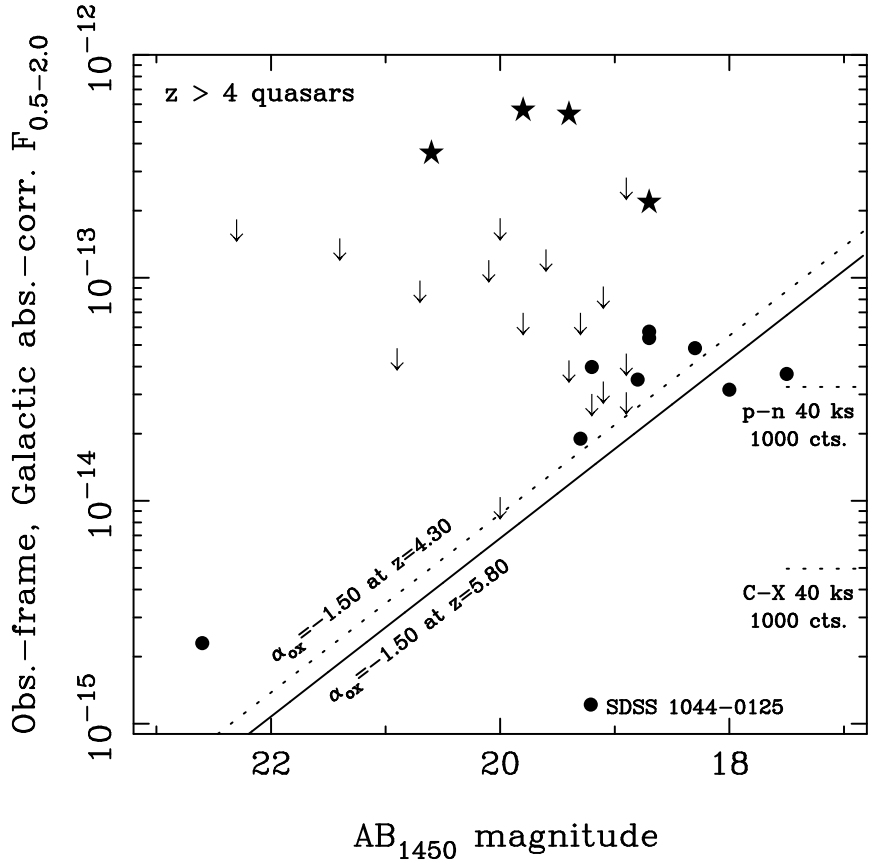}
}
\caption{AB$_{1450}$ magnitude versus observed-frame, Galactic 
absorption-corrected 0.5--2.0~keV 
flux for quasars at $z>4$ (updated from Kaspi et~al. 2000). The 
solid dots are X-ray detected RQQs, the stars are X-ray 
detected radio-loud blazars, and the arrows are X-ray upper limits.
The slanted lines show $\alpha_{\rm ox}=-1.50$ loci for
$z=4.30$ and $z=5.80$ (assuming the same X-ray and optical spectral
shapes as are used for \sdss\ in the text). Note that, compared to its
peers, \sdss\ is quite X-ray faint given its AB$_{1450}$ magnitude. 
The two horizontal dotted lines near the right edge of the plot 
show the observed-frame, Galactic absorption-corrected 0.5--2.0~keV 
fluxes required to gather 1000 counts (from 0.5--2.0 keV) in a typical 40~ks 
observation for the \xmm\ EPIC p-n and the planned \conx\ microcalorimeter; we have 
adopted $\Gamma=2$ and $N_{\rm H}=4.6\times 10^{20}$~cm$^{-2}$ when 
computing these values.
This plot is useful for practical work (e.g., observation planning)
and is fairly robust since it shows the directly measured fluxes
for objects with comparable redshifts.}
\end{figure}


\vspace*{-0.5truein}

\begin{figure}[t!]
\hbox{
\hspace*{0.6 in} \psfig{figure=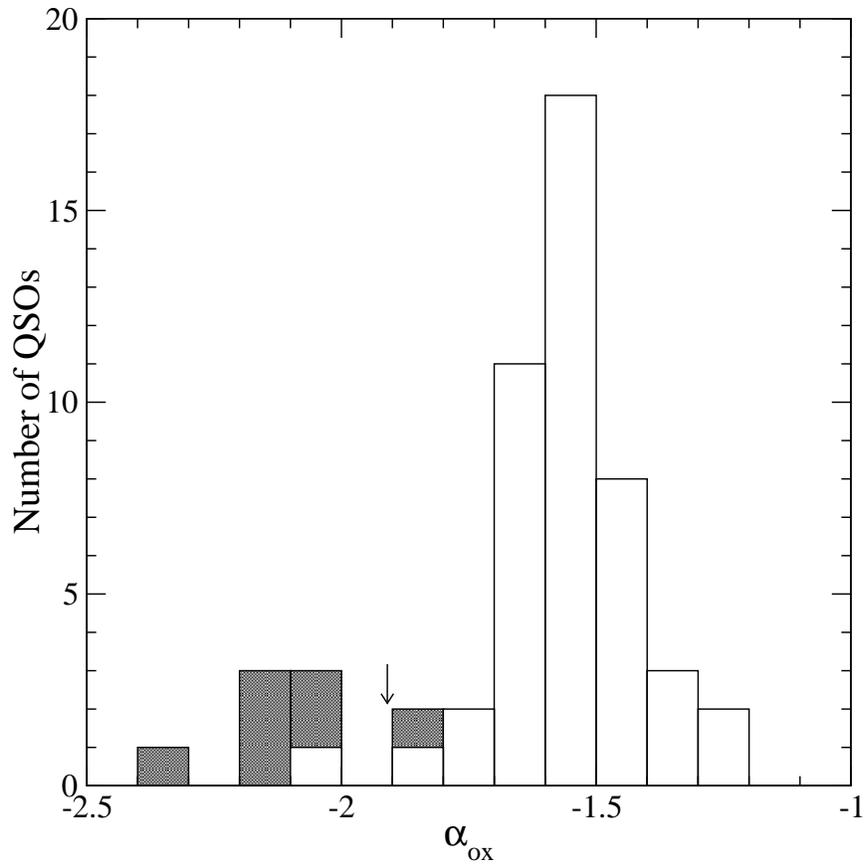,height=6.5truein,width=5.0truein,angle=0.}
}
\caption{Histogram of \aox\ for the 53 luminous ($M_{\rm V}<-22.27$) 
BQS RQQs. We have shaded the histogram for 7 of these RQQs, since there 
is strong evidence that their \aox\ values have been depressed 
by intrinsic absorption (see \S3.1.1). The downward pointing arrow 
marks the \aox\ value of \sdss.}
\end{figure}



\begin{figure}[t!]
\hbox{
\hspace*{0.6 in} \psfig{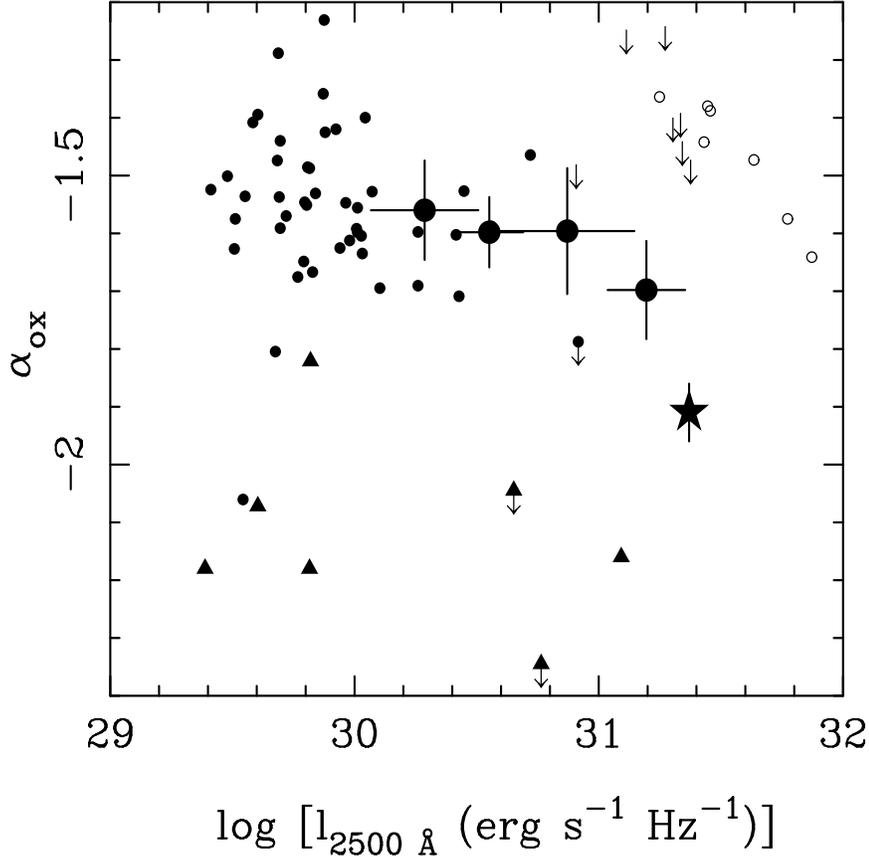}
}
\caption{\aox\ versus luminosity density at 2500~\AA\ for 
(1) the 7 luminous BQS RQQs from \S3.1.1 that have their $\alpha_{\rm ox}$
values strongly depressed by intrinsic absorption (solid triangles, two with
downward pointing arrows representing upper limits for the quasars
0043+039 and 1700+518), 
(2) the other 46 luminous BQS RQQs from \S3.1.1 (small solid dots, one with 
a downward pointing arrow representing an upper limit for the quasar 1259+593), 
(3) the LBQS RQQs from Figure~6b of Green et~al. (1995; large solid dots), and
(4) the optically selected $z>4$ RQQs from Kaspi et~al. (2000; open circles
for detections and plain downward pointing arrows for upper limits). 
The LBQS RQQ data points were derived using an X-ray counts stacking 
technique as described in Green et~al. (1995); from left to right, these 
data points represent 21, 21, 35 and 70 LQBS RQQs. 
\sdss\ is shown as the star; note its large negative value of \aox. 
The three small solid dots notably below the general trend 
(with $\alpha_{\rm ox}\leq -1.8$) are, from left to right, 
2214+139, 1552+085 and 1259+593 
(see \S3.1.1 and BLW for discussion).}
\end{figure}


\end{document}